\documentclass[aps,prl,preprint,showpacs,floatfix]{revtex4}
\usepackage{graphicx}
\begin{document}

\title{Nature of Charge Carriers in Disordered Organic Molecular Semiconductors} 

\author{Ajit Kumar Mahapatro and Subhasis Ghosh}

\affiliation{School of Physical Sciences, Jawaharlal Nehru University, New Delhi 110067, India}

\begin{abstract}
 
 Understanding the charge carrier transport in the disordered organic molecular semiconductors is a fascinating and still unresolved problem in modern condensed-matter physics, yet has an extremely important bearing on their application in ``plastic'' based field effect transistor, light emitting diodes and lasers. The most contentious    issue in this subject is the nature of the charge carriers i.e.   whether they are bare electrons or dressed electrons, known as polarons.   Here we show for the first time, by means of simple experiments that polaron moving in static disorder is responsible for charge carrier transport in disordered molecular semiconductors.

\end{abstract}


\maketitle

Since the discovery of ``plastic'' based transistor\cite{fg94,zb96} and light emitting devices\cite{cwt87,jhb90,nt96}, disordered molecular semiconductors are being widely investigated.   These materials are composed of different types organic molecules starting from very small molecules to very large biomolecules like proteins and DNA, held together loosely by   weak van der Waals type intermolecular force. The absence of long range order in these materials leads to the localization of the electronic wave function and  results transport of charge carrier  via thermally activated hops with  carrier mobility $\mu$ as a strong function of  temperature and electric field $F$ following  a universally\cite{dhd96} observed  Poole-Frenkel(PF) behavior $\mu(F,T)=\mu(0,T) \exp(\gamma \sqrt{F})$,  where  $\mu(0,T)$ is the  temperature dependent  zero field mobility and $\gamma$ is  field activation of the mobility and recently,  in light of  theoretical studies\cite{svn98,pep02,zgy00} we have shown\cite{akm03} another important feature i.e. (iv) charge carriers are spatially correlated in these materials. Considerable debate is going on the fundamental question like,  {\sl what is the nature of charge carriers in disordered molecular semiconductors}? Answer to this question   will be a key input for understanding the charge transport mechanism and the development of new materials and new devices based on these materials.   An indication of the difficulty in finding the nature of charge carriers and understanding the transport processes in these materials is evidenced by the conflicting viewpoints. At present, there are three viewpoints on the nature of charge carriers, which are   (i) dressed electrons(known as polarons) due to strong electron-phonon interaction and the temperature dependence of the $\mu$ arises due to polaron binding energy\cite{lbs90}; (ii) bare electrons and temperature dependence of $\mu$ is due to energetic and spatial disorder\cite{pmb91} and (iii) third viewpoint\cite{pep02,zgs00} is the combination of these two opposite viewpoints and the charge carriers are polarons moving in disorder and the temperature dependence of $\mu$ is due to both static disorder and polaron binding energy.  It has been shown\cite{pep02,lbs90} that polaron based models results unacceptable values of polaron binding energy and strong nearest neighbor intermolecular coupling if disorder is not included. As mentioned earlier, the intermolecular coupling in these disordered molecular materials is very weak and can't account for strong nearest neighbor coupling. If it is assumed that energetic and spatial disorder are responsible for thermal activation of $\mu$ and bare electrons, not polarons, are responsible for charge carrier transport, the probability of hopping is drastically reduced and again unreasonably strong intermolecular coupling is required to explain the experimental data.  Here, we show that there is a natural solution to this apparent paradox and polaron is the nature of charge carriers, emphasizing the important role of both disorder and electron-phonon interaction on the charge carrier transport in these disordered materials. 

We have chosen one of the most important\cite{zb96,gp98} molecular solid based on Copper phthalocyanine(CuPc) molecules because of their high chemical stability and ease to prepare the devices.  Details about the CuPc-based single layer devices are given in Ref.\cite{akm02}. Here, we report the experimental investigation on charge carrier transport in hole only devices based on metal/CuPc/metal structures. In this case, by properly choosing contacting metals, current injection and transport due to holes have been
investigated. Fig.1 shows the temperature dependent current-voltage(J-V) characteristics in  ITO/CuPc/Al structure.  At low bias, J-V characteristic follow the slope of two in log(J)-log(V) plot, but as the bias increases the slope   increases gradually to higher value due to space charge limited transport process\cite{akm01}, because the ionization potential of MePc(4.8eV)\cite{gp98} is close to the work function of ITO(4.75eV). A numerical workout\cite{akm01} has been used by solving Poission's equation $dF(x)/dx=ep(x)/\epsilon$,  current density relation  $J(x)=ep(x)\mu(x)F(x)$  and  the PF relation of carrier mobility  $\mu(F,T)=\mu(0,T)\exp(\gamma\sqrt{F})$, simultaneously. Here, $\epsilon$ is the dielectric constant of CuPc and $p(x)$ is the hole density at position $x$ and  the boundary condition is taken\cite{akm02,akm01} at the ITO/CuPc interface. The simulated results are shown as solid lines in  Fig.1 at different temperatures.  Inset of Fig.1 shows the temperature dependence of conductivity $\sigma(T)$ for CuPc. Linear dependence of $ln(\sigma)$ on $1/T^{1/4}$  validates the  variable range hopping\cite{nfm79}, $\sigma(T)=\sigma_0 exp\{-(T_0/T)^{1/4}\}$, where $\sigma_0$ is the temperature independent conductivity and $T_0$ is the characteristic temperature.  Hence the excellent  agreement of the simulated results with experimental data presented in Fig.1  establish the carrier transport process, which is thermally activated hopping with field dependent  mobility. 

There are basically three models to explain the charge carrier transport in these materials. In a very first approach,  Gill\cite{wdg72} attempted to describe the universally observed thermally activated and PF behavior of $\mu$ using a phenomenological non-Gaussian disorder model(NGDM), $\mu(T,F\rightarrow 0)\propto\exp(-\Delta/k_BT)$, where  $\Delta$ is the temperature independent activation energy. This model has been criticized\cite{svn98,akm03,hb93} for lacking a first principle explanation of PF behavior of $\mu$ and non-inclusion of another universal feature i.e. the Gaussian distribution of energy levels, $g(E)\propto\exp(-E^2/4\sigma^2)$, where $\sigma$ is the r.m.s deviation of hopping site energies.   Bassler  proposed\cite{hb93} the uncorrelated Gaussian disorder model(UGDM), which describes the carrier transport as a biased random walk among the dopant molecules with Gaussian-distributed random site energies due to positional and energetic disorder.   Although UGDM explains some features of experimental data and provides support for PF behavior of carrier mobility, several discrepancies\cite{akm03} emerge with uncorrelated description of Gaussian disorder model. Garstein and Conwell\cite{yng95} first showed that a spatially correlated potential is required for the description of PF behavior of mobility and subsequently Novikov {\sl et al.}\cite{svn98} have proposed a correlated Gaussian disorder model(CGDM) to describe  the charge carrier transport in these materials. 
Recently, we have shown\cite{akm03} that the CGDM successfully explains the experimental data emphasizing, whatever be the nature of charge carriers, charge transport occurs by correlated hopping among the molecular sites distributed in Gaussian density of states(GDOS) of highest occupied molecular orbital(HOMO) and/or lowest unoccupied molecular orbital(LUMO) in these disordered materials. 
In case of both Gaussian disorder models, the temperature dependence of zero field mobility is given by $\mu(T,F\rightarrow 0)\propto\exp(-A \sigma^2/k_B^2T^2)$, where $A$ is $2/3$ in UGDM\cite{hb93} and $3/5$ in CGDM\cite{svn98}.  
The Aerrhenious plots result an apparent activation energy $E_c$, which is $E_c=-k_B\{d(ln\mu)/d(1/T)\}=\Delta$ in  NGDM\cite{wdg72} and  $B \sigma^2/k_BT$,  in case of  Gaussian disorder models and $B$ is $8/9k_B$ in UGDM\cite{hb93} and $18/25k_B$ in CGDM\cite{svn98}. If we take the contributions both from the polaron binding energy $E_{pol}$, and from the static disorder $E_{dis}$,  $E_c$ should be $E_c=E_{pol}+E_{dis}$. Fig.2 shows the temperature dependence of zero field mobility and it is clear that both $ln(\mu)$ vs. $1/T$ and $ln(\mu)$ vs. $1/T^2$  result nonlinear Arrehenious plots and it is impossible to distinguish between $E_{pol}$ and $E_{dis}$ and their respective values from $E_c$.    Here, we show that the interplay between these two contributions on the carrier transport can be explored by studying the temperature dependence of $\mu$ over extended range of temperature and measuring atleast one contribution in E$_c$, independently. 
It is clear from Fig.2 that both,  NGDM  and UGDM or CGDM agree well with experimental data in higher temperature region($T\geq140K$) and gradual change in slope in the Arrehenious plots can be explained   either by invoking  the distribution of  $E_c$, or by distribution of $\sigma$,  but  the underlying assumption behind the  second proposition  is hard to justify. We have found that the  Gaussian distribution  of $E_c$ i.e. $\phi(E_c)=\exp[-(E_c-E_{c0})^2/4\sigma_c^2]$ and  $\mu(0,T)=\mu_0\int^\infty_0\phi(E_c)\exp(-E_c/k_BT)dE_c$  explain the data excellently(shown in Fig.3).  $E_{c0}$, which is the maximum activation energy  and $\sigma_c$,  are the center and width of the Gaussian distribution and found to be 540meV and    80meV,  respectively.  In this case,  the temperature dependence of mobility  follows $ln(\mu)\propto 1/k_BT$ instead of  $ln(\mu)\propto 1/(k_BT)^2$ according to  CGDM\cite{svn98}. The spatial correlation in CGDM\cite{svn98} has been   shown to be resulted from   the long-range interaction between charge carriers and permanent dipole moments of doped molecules in polymers and/or  host molecules. 
However, it has been pointed out\cite{zgy00} that the mechanism responsible for PF behavior in different conjugated polymers and molecules cannot be due to charge-dipole interaction,  because the  PF behavior of $\mu$ has been universally observed in several doped and undoped organic semiconductors based on polymers or molecules with or without dipole moment.  Recently, Yu {\sl et. al.}\cite{zgy00} have shown using first principle quantum chemical calculation that the thermal fluctuations in the molecular geometry can lead to spatial correlation through intermolecular restoring force and different temperature dependence in this case arises due to the fact that energetic disorder is temperature independent in CGDM, whereas it increases with temperature in molecular geometry fluctuation model\cite{zgy00,zgy01} resulting $ln(\mu)\propto 1/k_BT$ temperature dependence. We have used highly symmetric molecules with negligible dipole moment for our investigations and we find it hard to justify strong long-range charge-dipole interaction to be the origin of spatial correlation.

Fig.4 shows the absorption spectra of CuPc. The characteristic two-humped spectra are known as Q-band({\sl 14}) in CuPc. The absorption co-efficient $\alpha(E)$ for a given photon energy E is proportional to the probability $P_{if}$ for the transition from initial state $i$(in HOMO) to the final state $f$(in LUMO) and the density of the initial and final state and given by, $\alpha(E) \propto \Sigma P_{if}${\large $\rho$}$_i${\large $\rho$}$_f$, where  {\large $\rho$}$_i$ and {\large $\rho$}$_f$ are the density of states in HOMO and LUMO of CuPc, respectively.
There are two important parameters involved in the absorption spectra. First is the  center of the peak at around 2eV, which is the energetic separation between maxima of HOMO and LUMO and second is the full width at half maximum of   about 460meV, which should be the maximum energetic separation between two hopping sites within the intrinsic GDOS, as shown schematically in Fig.4. 
Hence, the maximum contribution in the activation energy from static disorder will be around 460meV. We have also found that the 540meV is the maximum activation energy($E_{c0}$), which has both contributions  from static disorder and polaron binding energy. Combining these two electrical and optical measurements, we found   80 meV(=540-460meV) is the   polaron binding energy. If we include the lower and upper limits  of activation energy due to static disorder, which are 500meV and 580meV respectively, the polaron binding energy varies between 40 to 120meV.   These observations are consistent with the third viewpoint i.e. disordered polaron moving in a correlated energy landscape is responsible for charge transport in  disordered organic molecular semiconductors.  These experimental findings   settle an extremely important issue regarding the nature   of charge carriers  and  have shown conclusively that polarons  moving among static disorders are responsible for charge carrier transport in disorder molecular semiconductors.

\newpage

\noindent {\large \bf Figure captions}

\vspace{0.15in}

\begin{description}

\item[1] J-V characteristics of    100nm  CuPc based single layer device  with ITO cathode contact and Al anode contact  at different temperatures starting at 300K and then at the interval of 30K.  Experimental data are shown by the symbols and solid lines represent the  simulated data. Inset shows the log of conductivity of CuPc vs. 1/T$^{1/4}$. Empty circles are experimental data and solid line is fit with straight line.

\item[2] Temperature dependence of zero field mobility against $1/T$(a) and $1/T^2$(b). Solid lines are linear fit to data, which results in (a) $\mu_0=2\times 20^{-4}cm^2/Vsec, \; \Delta=450meV$ and in (b) $\mu_0=2\times 10^{-8}cm^2/Vsec, \: \sigma=85meV$.

\item[3] Theoretical fit to temperature dependence of zero field mobility in CuPc for two different thicknesses  using Gaussian distribution of activation energy, which is shown in inset.

\item[4] Absorption spectra of Cu-Pc for two different thicknesses at room temperature.  The spectral characteristics and width of the spectra do not depend on the thickness of the sample. Inset shows the hopping sites and hopping processes inside the intrinsic density of states of disordered molecular semiconductors.
 
\end{description}

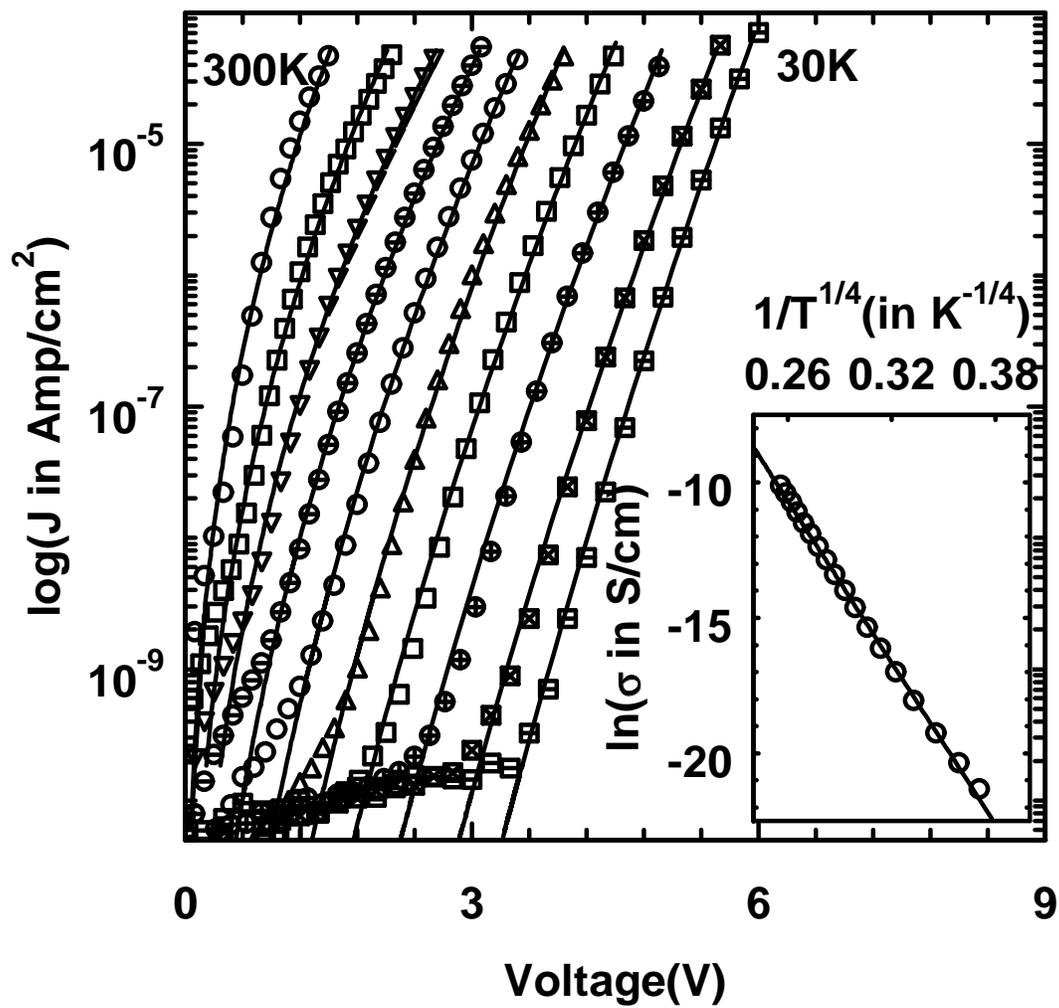
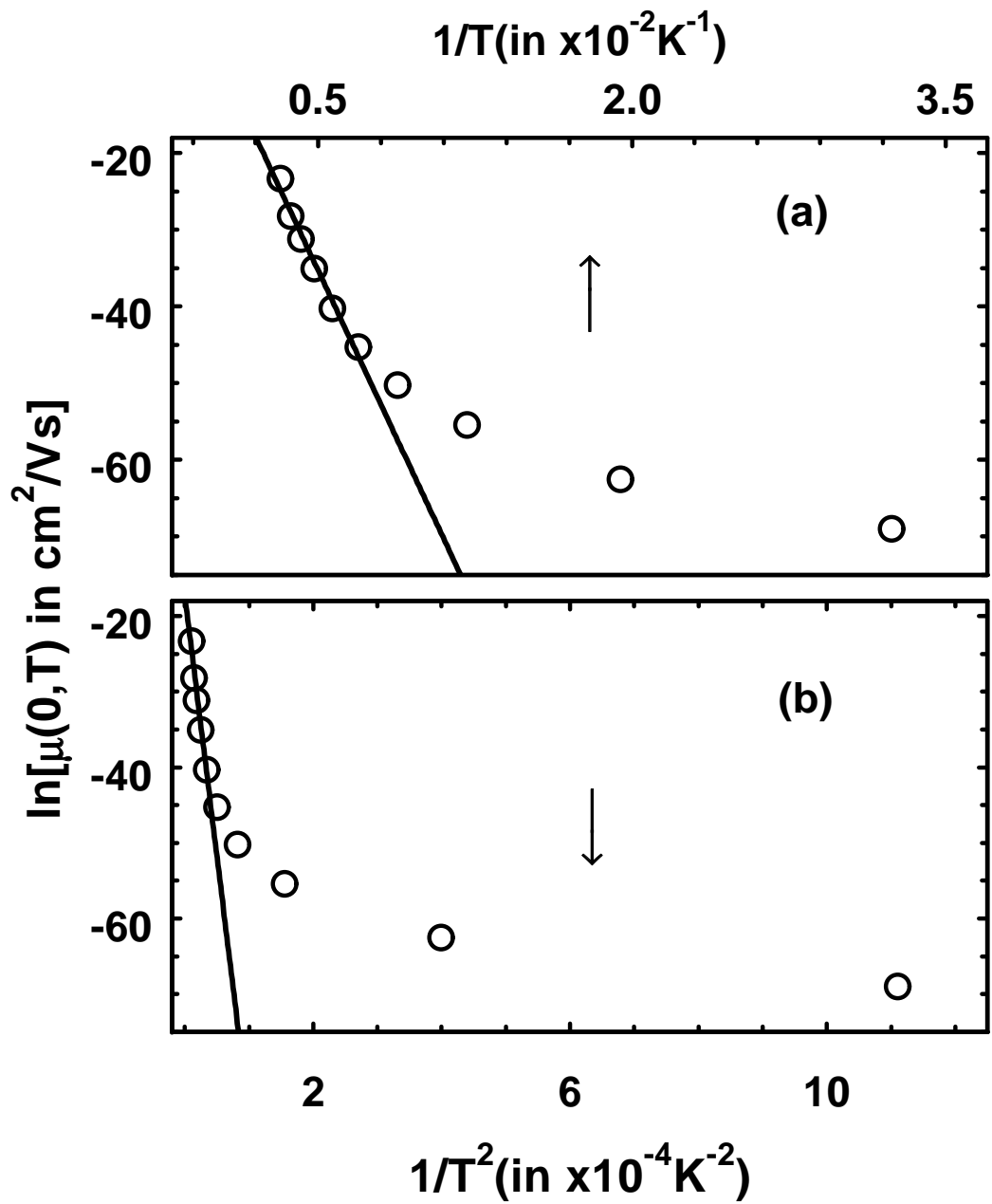
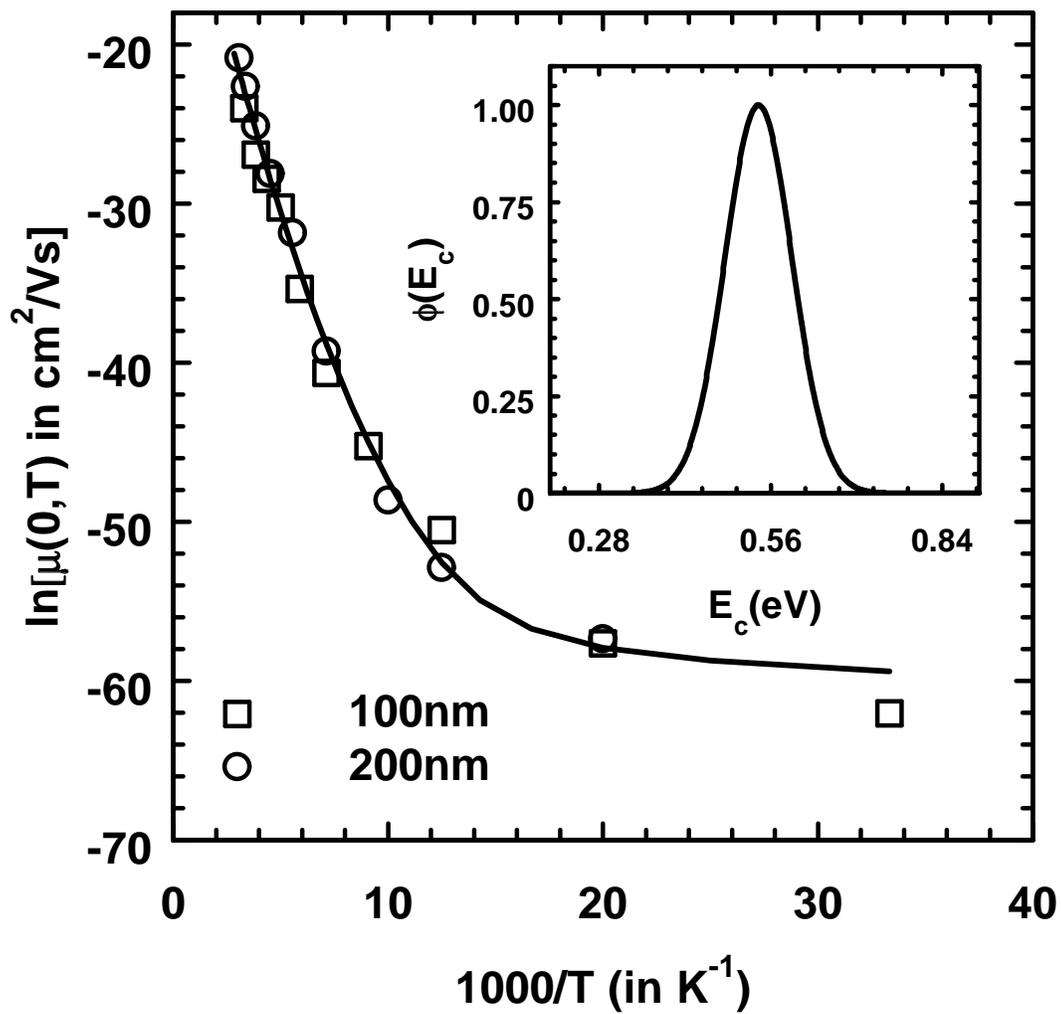
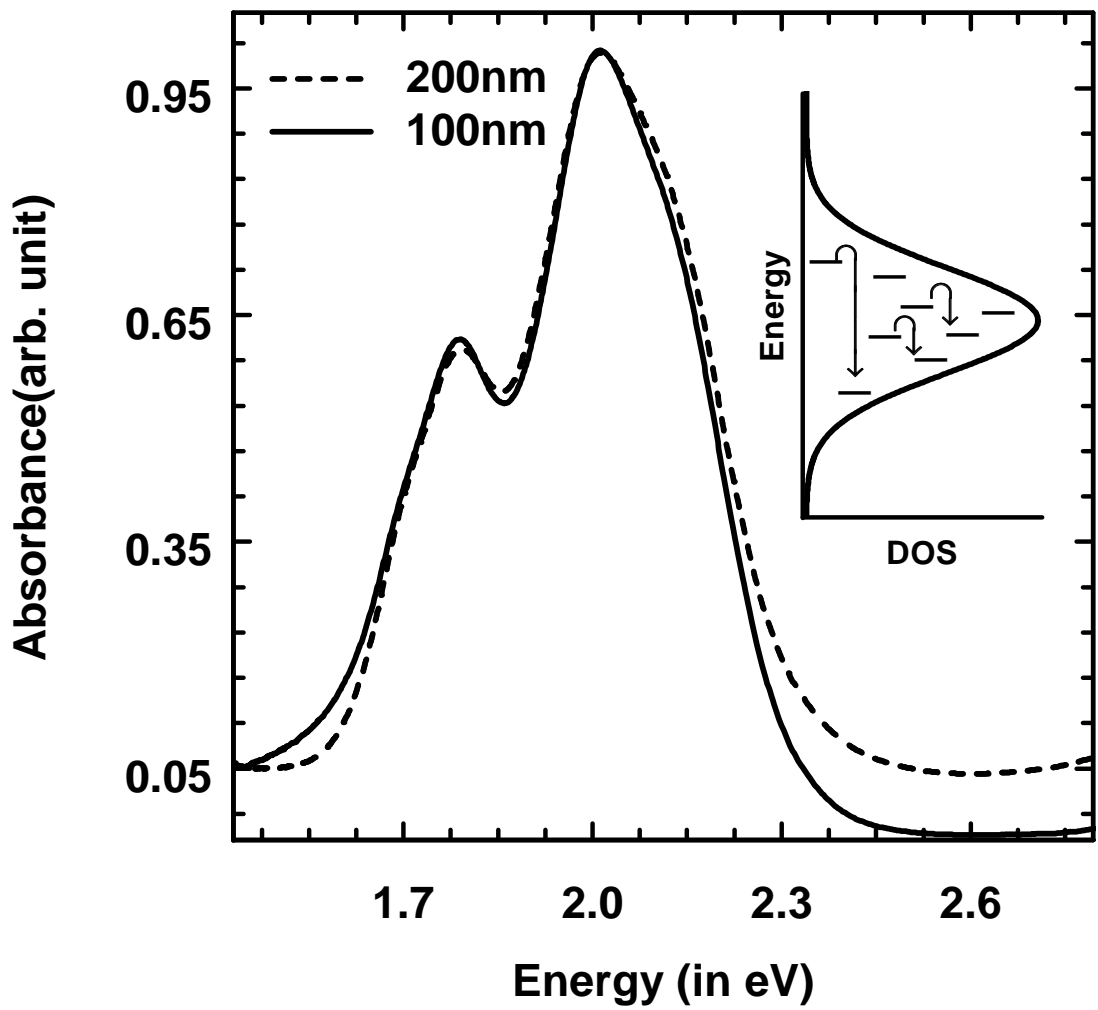
\end{document}